\documentclass[usenatbib,longtabl]{mn2e}
\input psfig.sty

\title[The new SDSS WDMS binaries]
{White-dwarf+main-sequence binaries identified from the tenth data
release of Sloan Digital Sky Survey}

\author[Li et al.]
{Lifang Li$^{1,2}$ \thanks{E-mail:
llf@ynao.ac.cn},Xiaobo Gong$^{1,2,3}$,Fenghui Zhang$^{1,2}$,Quanwang Han$^{1,2,3}$, and Xiaoyang Kong$^{1,2,3}$\\
$^1$Yunnan Observatories, Chinese Academy of Sciences, P.O.Box
110,
Kunming, Yunnan Province 650011, China\\
$^2$Key Laboratory for the Structure and Evolution of Celestial
Objects, Chinese Academy of Sciences, P.O. Box 110, Kunming,\\
Yunnan Province, 650011, China\\
$^3$University of the Chinese Academy of Sciences, Beijing,
100049, China}

\begin{document}

\date{Accepted yy mm dd. Received yy, mm, dd; in original form 2014 December 1}

\pagerange{\pageref{firstpage}--\pageref{lastpage}} \pubyear{2014}

\maketitle

\label{firstpage}

\begin{abstract}
We have presented 309 new white-dwarf (WD) + main-sequence (MS)
star binaries identified from the Sloan Digital Sky Survey (SDSS)
Tenth Data Release (DR10). The majority of them consist of a white
dwarf and a low-mass secondary (typically M dwarf) companion. The
SDSS spectra of the newly found WDMS binaries with a DA/DB white
dwarf and an M/late-K dwarf companion are analyzed based on a
spectral decomposition/fitting method. White dwarf effective
temperatures, surface gravities and masses together with the
secondary star spectral types are obtained, and the stellar
parameters of DA WDs with $T_{\rm eff}\la 14,000$ K are revised to
the results in the case of 3D model atmosphere. Two independent
distance estimates are derived from the flux-scaling factors
between the WDMS SDSS spectra and the white dwarf and M-dwarf
model spectra. It is found that about more than 20 per cent of the
newly found WDMS binaries show a significant discrepancy between
the two distance estimates. This might be caused by the effects of
M-dwarf stellar activity or irradiation of the M dwarf companions
by the white dwarf. The stellar parameter distributions are used
to investigate the global properties of newly found WDMS binaries,
the results shown in this work are consistent with those derived
by previous investigators. Some WDMS binaries have been observed
more than one ($2-4$) times by SDSS, it is found that four of them
exhibit not only Hydrogen emission in all observable Balmer series
lines in addition to He I line Balmer lines but also the rapid
variation in the radial velocities of the components in these
binaries. This suggests that they should be the post common
envelope binaries (PCEBs) with a short orbital period. These young
PCEBs are very important for limiting the results of common
envelope phase of the binary evolution.
\end{abstract}


\section{Introduction}
White-dwarf + main-sequence (WDMS) binaries are very important
celestial objects since they constitute a common final state
objects of stellar evolution, a WD, and a MS star. They are the
potential progenitors for most cataclysmic variables and perhaps
type Ia supernova (SN Ia) and are therefore a subject which can
inspire the keen interest of many investigators \citep{Sil06,
Heller09}. The WDMS binaries are compact binaries containing a WD
component whose progenitor star should have undergone a giant star
(usually during red giant branch  or a asymptotic giant branch)
phase as a single star \citep{Reb10}, the progenitor binaries of
the WDMS binaries should be the wide initial ones. It is well
known that the orbital separation of all types of close compact
binaries, such as cataclysmic variables (CVs), X-ray binaries and
double degenerate binaries are much shorter than the radius ($\sim
100 R_{\rm \odot}$) of a giant star \citep{max09}, implying that a
large amount of angular momentum and orbital energy of their
progenitor binaries have been lost by an unusual physical
mechanism so called a common envelope (CE) evolution phase
\citep{pac76,web84,web08,Iben86,Iben93}. Up to now, the CE phase
is believed to be a result of dynamically unstable mass transfer
from a giant star to its MS companion \citep{pac76,web84,hje89}.
At the beginning of the CE evolution, the secondary star starts a
spiral-in process, then the CE is gradually ejected from the
binary system owing to the friction induced by asynchronous
rotation in CE. When the CE has been completely ejected from the
system at the end of CE evolution, a post common envelope binary
(PCEB), which consists of a compact object and a companion star,
appears.

The CE evolution phase plays a central role in many evolutionary
pathways leading to the formation of compact objects in the short
period binary systems \citep{Taa00,Taa10}, such X-ray binaries as
natural laboratories for general relativity, CVs and double
degenerate WD binaries as the potential progenitors of SN Ia,
together with double degenerate neutron star binaries as the
progenitors of short gamma-ray burst \citep{reb12}. Although the
main features during the CE evolution phase have been depicted by
\citet{pac76} more than 30 years ago, and a lot of works on the
formation theory of PCEBs have been carried out by various
investigators \citep[e.g.][]{Iben93,Davis08,Taa00,Pos06,Pol10},
however, owing to the complex physical processes occurred in the
CE phase and lack of the observational constraints, our
understanding about CE phase is still very poor
\citep{reb07,reb12}. Since the detached PCEBs have not been
severely contaminated by mass transfer after they formed through a
CE evolution, they are thought to be the most suitable objects for
the constraints of the CE evolution result. That is to say, they
are almost the real materials produced by the CE evolution except
that the size of their orbit has been decreased by AM loss owing
to magnetic stellar wind and/or gravitational wave radiation.
Among all PCEBs, those composed of a WD and a MS star represent
the most promising population for deriving such observational
constraints, as they are essentially most common population of
PCEBs and their physical parameters can be derived from
observations through 2-8m telescopes \citep{Neb11}.

One of the many interesting by-products of the Sloan Digital Sky
Survey \citep[SDSS;][]{York00,Sto02,Ade08,Aba09} is the
identification of a large number of unresolved, close binary
systems. In particular, the five-color ($u,g,r,i$ photometry and
$z$) and spectra covering almost the entire optical wavelength
range permit to easy identification of binaries contained a blue
white dwarf (blue) and a red low-mass M dwarf secondary
\citep{Sil06}. Therefore, the SDSS has been thought to be very
efficient in searching for WDMS binaries. Up to now, more than
2,500 WDMS binaries have been discovered in SDSS survey
\citep{Sil06,Sil07,Ade07,Heller09,Liu12,reb12,Reb13,Li14} since
\citet{Ray03} and \citet{Sch03} first attempted to investigate the
WDMS binaries in the SDSS, however the orbital periods of only
about 200 WDMS binaries have been determined by the various
investigators through the radial velocity observations, and is
clearly very insufficient to limit the results of the CE
evolution. This suggests that the orbital periods of of most WDMS
binaries are difficultly determined by observations due to
insufficient telescope time for their radial velocity
observations. Therefore, it is necessary to find more WDMS
binaries from the new data released by SDSS or by other spectral
survey telescope, such as LAMOST \citep[also called GuoShouJing
Telescope,][]{Ren13,Wei12}.

In this work, we present 309 new WDMS binaries, which are mainly
composed of a white dwarf and an M dwarf, newly identified from
SDSS (DR10) based on a color-selection criteria \citep{Li14}
originally developed by \citet{Liu12} and a criteria developed by
Gong et al. (2014, in preparation). We have analyzed their SDSS
spectra by using a method proposed by \citet{Heller09}, then
derived the fundamental physical parameters for the two spectral
components in each WDMS binary which are composed of a DA/DB white
dwarf and an M-dwarf. At last, based on the stellar parameter
distributions, the general properties of the WDMS binaries
identified from SDSS by us are discussed.

\section{WDMS binaries identified from SDSS DR10}

Based on a color-selection criteria \citep{Li14} originally
developed by \citet{Liu12} and a color-selection criteria for
searching for WD stars from LAMOST and SDSS (Gong et al. 2014, in
preparation), we identified 2198 WDMS binaries from SDSS DR10, and
1889 WDMS binaries of them had been identified by the previous
investigators \citep{Sil07,Heller09,Liu12,Wei12,Reb13,Li14}.
Therefore, 309 new WDMS binaries are identified from SDSS DR10. In
these new binaries, 220 binary systems contain a DA WD component,
32 systems contain a DB WD component, and 57 systems contain a WD
component with the other spectral types or an unknown spectral
type because of a low SN. The infrared counterparts of some new
WDMS binaries are obtained by cross-match (in a radius of 5
arc-seconds) with the Two Micron All Sky Survey
\citep[2MASS:][]{Skr06}, the UKIRT Infrared Sky Survey
\citep[UKIDSS:][]{Dye06,Law07,War07}, and the Wide-field Infrared
Survey Explorer \citep[WISE:][]{Wri10}. The object name,
coordinates (in degrees) and SDSS $ugriz$, UKIDSS $yjhk$, 2MASS
$JHK$ and WISE $w_1, w_2$ magnitudes of 311 new WDMS binaries are
listed in Table 1.

\section{Stellar parameters}

\subsection{Stellar parameters of new WDMS binaries containing a DA/DB WD companion}

Based on a method, which can remove a mutual dependence of the
scaling factors and the effects of identifying a local $\chi^2$
minimum, proposed by \citet{Heller09}, we decomposed/fitted the
observed spectra of the new WDMS binaries (with a DA/DB WD
component) identified from SDSS DR10. In the course of the fitting
of their spectra, the theoretical spectral grid of the DA WDs is
the same as \citet{Liu12} and \citet{Li14} who employed a
theoretical spectral grid developed by \citet{Koes10}. In order to
obtain the stellar parameters of some WDMS binaries with a DB WD
star component, we employ a theoretical spectral grid of DB WDs
also developed by \citet{Koes10}, in which the surface gravity,
${\rm log}g_{_{\rm WD}}$, covers a range from 7.0 to 9.0 in steps
of 0.25, and the effective temperature, $T_{\rm WD}$, covers a
range from 8,000 K to 40,000 K, in steps of 500 K for $T_{\rm
eff}^{\rm WD}\la 10,000$ K, 1,000 K for $10,000\ {\rm K}\la T_{\rm
eff}^{\rm WD}\la 20,000\ {\rm K}$, and 2,000 K for $T_{\rm
eff}^{\rm WD}\ga 20,000\ {\rm K}$. With the help of these
theoretical spectral grids, we can obtain the surface gravity and
effective temperature of DA/DB WDs in WDMS binaries identified
from SDSS DR10. The parameter spaces of the surface gravity and
effective temperature for the M dwarfs are the same as those used
in \citet{Li14}. Since the determination of the M dwarf
metallicity is easily affected by the signal-to-noise ratio (SN)
of the observed spectra of the WDMS binaries which should be in
the neighborhood of the sun \citep{Li14}, the metallicity of the M
dwarfs is set to be the solar abundance, i.e. $[{\rm Fe/H}]=0$, in
this work, then the template spectra can be produced by the second
version of BaSeL standard library \citep{Lej97,Lej98}. In
addition, a few new WDMS binaries found from SDSS DR10 with a
K-dwarf star companion. In the course of the spectral fitting of
these WDMS binaries, the parameter spaces of the effective
temperature, surface gravity, mass and radius of the K-dwarf stars
are produced according to the theoretical models given by
\citep{Bar98}. The template spectra for K dwarfs are also given by
the BaSeL standard library. On the basis of the best fitting
result of the observed spectrum, we can obtain the effective
temperature and surface gravity of the MS star companion in each
WDMS binary. The results of the spectral analysis of three typical
WDMS binaries [SDSS J163033.26+393714.0 (containing a DA WD star
and an M dwarf companion), SDSS J111615.65+251716.1 (containing a
DA WD star and a K-type star companion) and SDSS
J103334.76+131159.6 (containing a DB WD star and an M dwarf
companion)] are shown in Fig. 1. The effective temperature and
surface gravity determined for the two components in WDMS binaries
through spectral analysis are presented in Table 2.

\begin{figure}
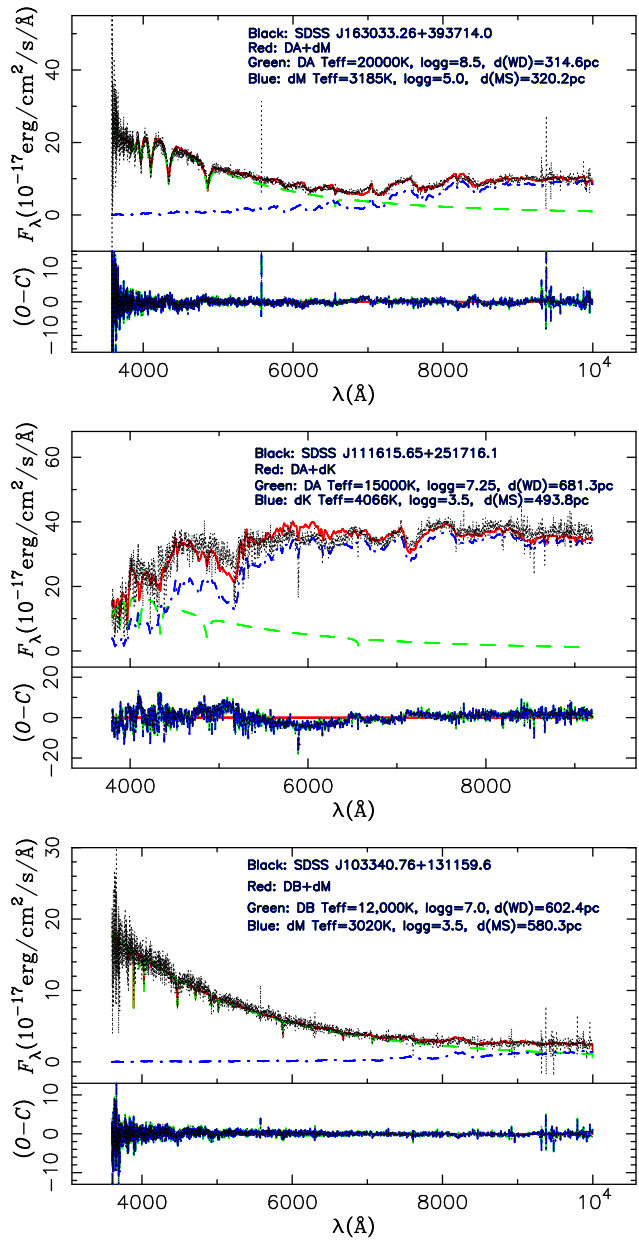

\centerline{\psfig{figure=Fig1.ps,width=8.5cm,height=5.5cm,bbllx=527pt,bblly=116pt,bburx=181pt,bbury=632pt,angle=270}}
\centerline{\psfig{figure=Fig2.ps,width=8.5cm,height=5.5cm,bbllx=527pt,bblly=116pt,bburx=181pt,bbury=632pt,angle=270}}
\centerline{\psfig{figure=Fig3.ps,width=8.5cm,height=5.5cm,bbllx=527pt,bblly=116pt,bburx=181pt,bbury=632pt,angle=270}}
\caption{Two-component fit to the spectra of the SDSS WDMS
binaries. Shown are examples for objects (DA+dM: SDSS
J163033.26+393714.0, DA+dK: SDSS J111615.65+251716.1, and DB+dM:
SDSS J103340.76+131159.6). In each panel, the dotted line
corresponds to the observed flux, the dashed line represents the
WD model flux, the dot-dashed line represents the main sequence
star model flux, and the solid line represents the total DA+MS
model flux. $O-C$ represents the residuals between the observation
flux and model flux.}
\label{fig1}
\end{figure}

Based on the spectral type of the M dwarf and the empirical
relations of $Sp-M$ and $Sp-R$ \citep{reb07}, the mass and the
radius of the M dwarf in each WDMS binary can be derived. For the
WDs in WDMS binaries, their parameters $T_{\rm eff}^{\rm WD}$ and
${\rm log}g_{_{\rm WD}}$ are determined from the best spectral
fitting based on results of a 1D model atmosphere \citep{Koes10}
at first, then they are revised to the results of a 3D model
atmosphere \citep{Tremblay13} for DA WDs with $T_{\rm eff}\la
1,4000\ {\rm K}$. The masses of the WDs can be determined
according to the newly updated cooling models for DA/DB WDs in
\citet{Ber95}, and their radii can be obtained through a recent
mass-radius relation presented in \citet{Hol12}. In order to
determine the cooling age of the WD components in the WDMS
binaries, we have made a distinction between WD stars with a
He-core and WD stars with a CO-core on the basis of their masses
derived from the spectral analysis as \citet{Li14} at first. The
WD stars with $M_{\rm WD}\ga 0.5 M_{\rm\odot}$ are defined to be
the CO WDs and the others are regarded as He WDs, although we
cannot conform that which type of WD with $M_{{\rm
WD}}\sim0.48-0.51 M_{\rm\odot}$ is most probable
\citep{hur02,zor11}. The cooling age of the He WDs is estimated by
interpolating cooling models of \citet{Alt97}, and the cooling age
of the CO WDs is estimated through a interpolation method
according to the cooling models of an updated version of
\citet{Ber95}. The values of these stellar parameters derived from
the results of the best spectral fitting are also listed in Table
2.

As \citet{Li14}, the two independent distances for the WD and
M-dwarf companions in WDMS binaries from the earth are estimated
on the basis of the best-fitting flux scales [$a$ and $b$ as
described in \citet{Liu12} and \citet{Heller09}] by the following
equations,

\begin{equation}
d_{_{\rm WD}}{\rm[pc]}=\frac{R_{\rm WD}{\rm [R_{\odot}]}}{1{\rm pc
}}\sqrt{\frac{\pi}{a}},
\end{equation}
and
\begin{equation}
d_{\rm M}{\rm[pc]}=\frac{R_{\rm M}{\rm [R_{\odot}]}}{1{\rm pc
}}\sqrt{\frac{\pi}{b}},
\end{equation}
where $R_{\rm M}$ is the radii of the MS components in WDMS
binaries, $d_{\rm M}$ is the distance of the MS components from
the earth, and the meaning of the other physical quantities in
eqs. (1) and (2) is the same as that described in \citet{Li14},
then a parameter $C$ used to describe the difference between
$d_{\rm WD}$ and $d_{\rm M}$ proposed by \citet{Heller09} is
written as the following:
\begin{equation}
C=\sqrt{2}\frac{d_{_{\rm WD}}-d_{\rm M}}{d_{_{\rm WD}}+d_{\rm M}},
\end{equation}
two independent distances for the WD and M dwarf from the Earth,
together with the parameter $C$, are also presented in Table 2.

\begin{figure}
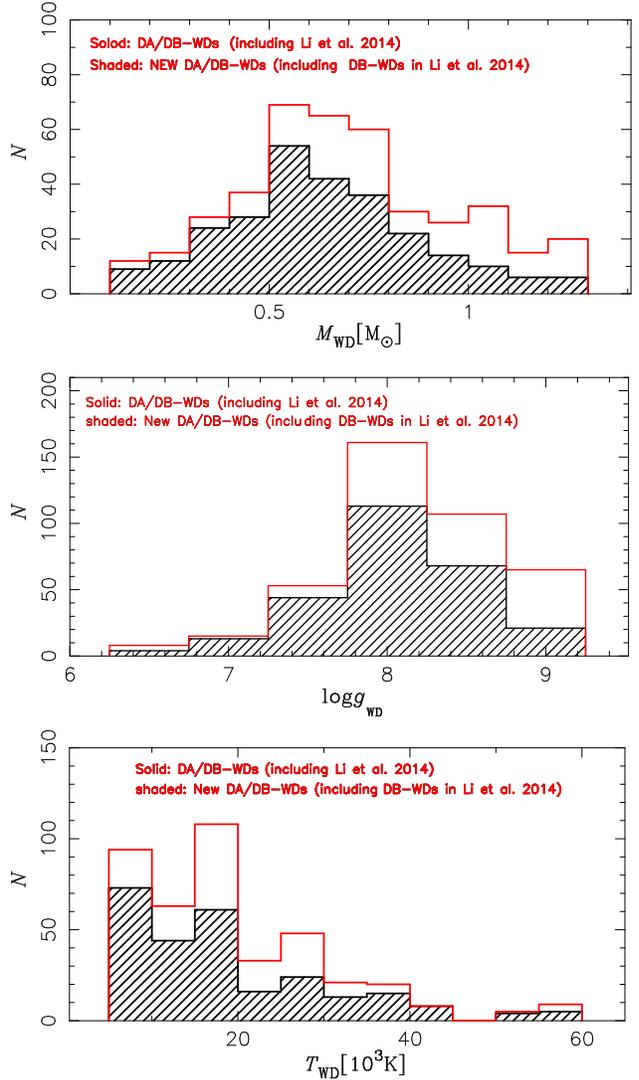

\centerline{\psfig{figure=Fig4.ps,width=8.5cm,height=4.8cm,bbllx=531pt,bblly=109pt,bburx=234pt,bbury=628pt,angle=270}}
\centerline{\psfig{figure=Fig5.ps,width=8.5cm,height=4.8cm,bbllx=531pt,bblly=109pt,bburx=239pt,bbury=630pt,angle=270}}
\centerline{\psfig{figure=Fig6.ps,width=8.5cm,height=4.8cm,bbllx=531pt,bblly=109pt,bburx=233pt,bbury=633pt,angle=270}}
\caption{White dwarf mass, ${\rm log} g_{_{\rm WD}}$ and effective
temperature distributions obtained from WDMS binaries identified
from SDSS. Solid line represents the parameter distributions of
all DA/DB-WDs identified by us from DR9 and DR10, and the shaded
zone represents the new DA/DB-WDs in the WDMS binaries newly found
from DR10 and in the WDMS binaries containing a DB WD component
found by \citet{Li14}.}
\label{fig2}
\end{figure}

\subsection{Distributions of stellar parameters of new WDMS binaries}

The stellar parameters of 205 new WDMS binaries with a DA WD
component and 32 new WDMS binaries with a DB WD component
identified from SDSS DR10 have been determined in Section 3.1.
Their global properties of the WDMS binaries are investigated
here. Fig. 2 shows the distributions of the mass $M_{\rm WD}$,
surface gravity ${\rm log}g_{_{\rm WD}}$ and effective temperature
$T_{\rm WD}$ of the DA/DB white dwarfs in 396 WDMS binaries [also
including 143 DA+MS binaries and 16 DB+MS binaries found from SDSS
DR9 by \citet{Li14}, solid line] and in 253 WDMS binaries with a
DA/DB WD component (the stellar parameters derived in this work,
shaded zone). As seen from the bottom panel of Fig. 2, the
effective temperature of the white dwarfs in most WDMS binaries is
located in a region between 6,000 K and 25,000 K, only about 25
per cent of the white dwarfs in WDMS binaries have an effective
temperature higher than 25,000 K in both cases. This should be
caused by a rapid cooling velocity for the white dwarfs with a
relatively high effective temperature \citep{Li14}. Meanwhile, it
is found in the intermediate panels of Fig. 2 that an average
surface gravity of the WDs is of about $10^{8}{\rm cm/s^2}$ and
the surface gravitation of most WDs is larger than 7.75. These
results are consistent with those derived by \citet{Liu12}. From
the top panel of Fig. 2, we can find that the mass of WDs in WDMS
binaries peaks at $\sim 0.5 M_{\rm\odot}$, and the average mass is
of $0.632 M_{\rm \odot}$ and $0.692 M_{\rm \odot}$ for the DA/DB
WDs in new WDMS binaries and in all WDMS binaries, respectively.
This is in good agreement with the results derived by the previous
investigators \citep[e.g.][]{Kepler14,reb07,reb12}.

\begin{figure}
\centerline{\psfig{figure=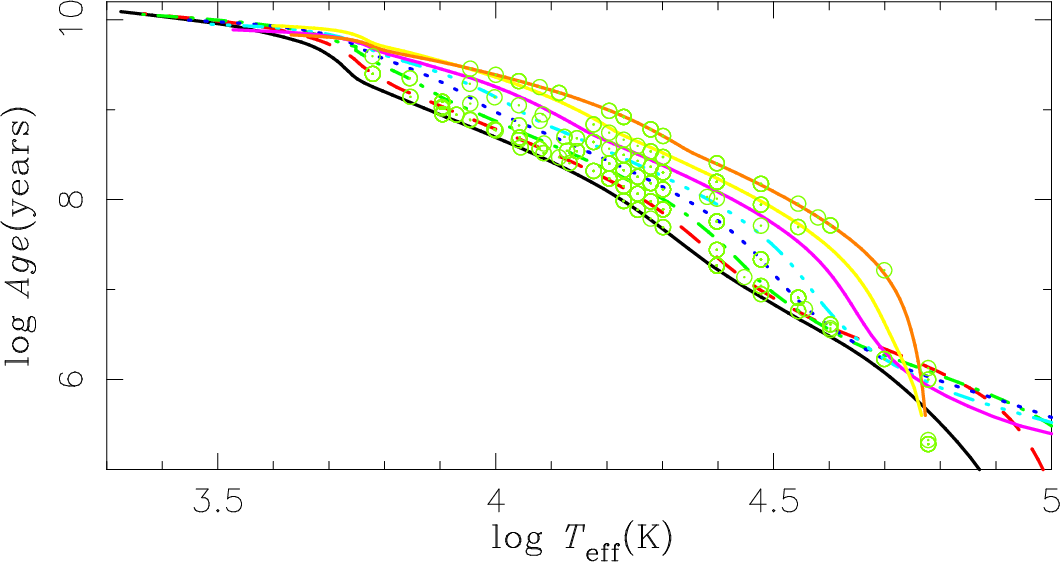,width=8.5cm,height=4.8cm,bbllx=525pt,bblly=109pt,bburx=246pt,bbury=628pt,angle=270}}
\centerline{\psfig{figure=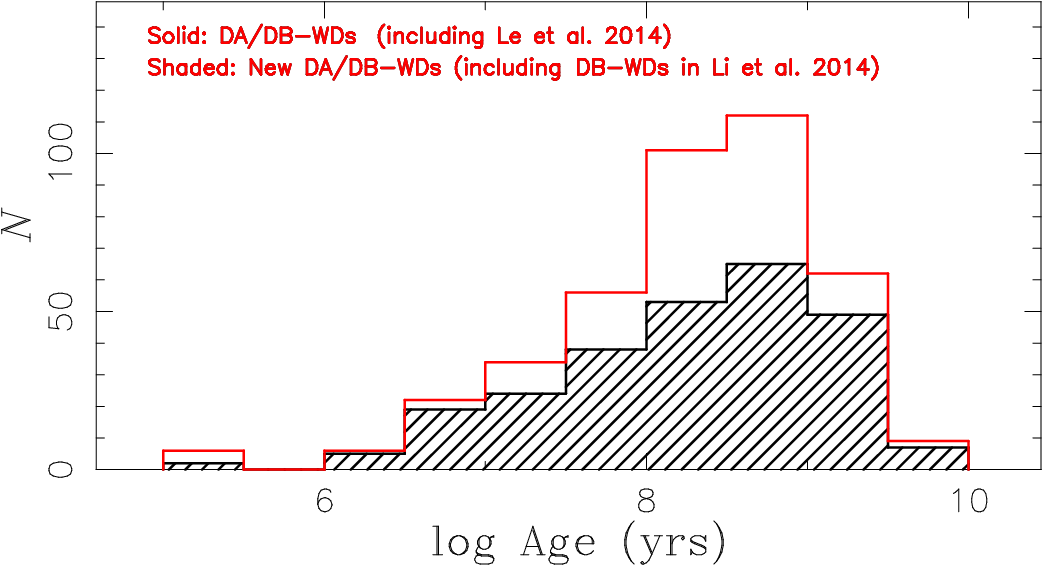,width=8.5cm,height=4.8cm,bbllx=525pt,bblly=109pt,bburx=246pt,bbury=628pt,angle=270}}
\caption{Top panel: The determination of the cooling age of CO
white dwarfs in WDMS binaries newly identified from SDSS DR10, The
lines represent the cooling models ($M_{\rm
WD}=0.5,0.6,...,1.2M_{\rm\odot}$) of \citet{Ber95}, the symbol
$\odot$ represents the observations. Bottom panel: The solid line
represents the distribution of the cooling ages of DA/DB-WDs in
all WDMS binaries found from DR9 and DR10, and the shaded zone
denotes that in those newly identified from SDSS DR10 and the WDMS
binaries containing a DB WDs found by \citet{Li14}.}
\label{fig3}
\end{figure}

The cooling age of the DA/DB WD components in our identified WDMS
binaries has been estimated by interpolation based on the cooling
models of He WDs \citep{Alt97} and CO WDs \citep{Ber95}. The
cooling age estimates of the white dwarfs in our WDMS binary
sample identified from SDSS DR9 and DR10 are shown in Fig. 3. The
cooling tracks of CO WD models with different masses (0.5, 0.6,
0.7, 0.8, 0.9, 1.0, 1.1 and 1.2 $M_{\rm\odot}$, repectively),
together with the estimates of cooling ages of only CO white
dwarfs, are shown in the top panel of Fig. 3. The bottom panel of
Fig. 3 shows the cooling age distributions of DA/DB WDs (solid
line) in 396 WDMS binaries found by us and in new WDMS binaries,
together with 16 WDMS binaries found by us from SDSS DR9 (shaded
zone), respectively. As seen from the bottom panel of Fig. 3, the
cooling age of the WDs in our WDMS binaries covers a region from
$10^5$ to $10^{10}$ yrs in both cases, and the cooling age of the
majority of DA/DB white dwarfs is larger than $10^8$ yrs. In
addition, the average cooling age of the WDs in our WDMS sample is
of about $2.0\times 10^8$ yrs. This suggests that the young DA
white dwarfs is very rare. This is also a result of the rapid
cooling velocity of the white dwarfs with a high effective
temperature. This is consistent with the result that the effective
temperature of most WDs are between 6,000 and 25,000K as shown in
Fig. 2.

\begin{figure}
\centerline{\psfig{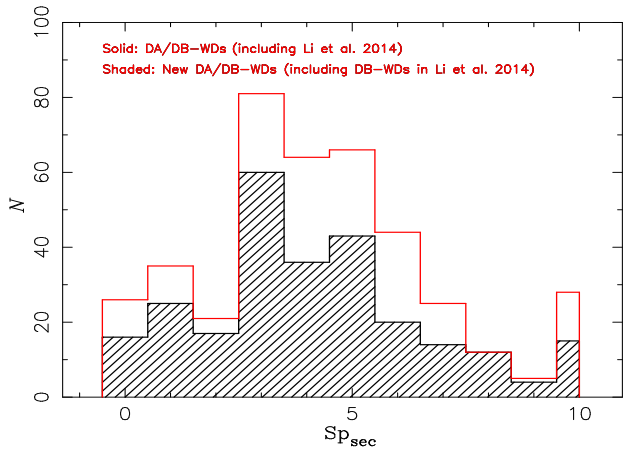}}
\caption{The distributions of the spectral types of the M Dwarfs
in WDMS binaries with a DA or DB WD component found from SDSS.
Solid line represents the spectral type distribution of M dwarfs
in all WDMS binaries, and the shaded zone represents that of M
dwarfs in WDMS binaries newly identified from DR10, together with
DB+dM binaries found by \citet{Li14}.}
\label{fig4}
\end{figure}

On the basis of the best fittings to their SDSS observed spectra,
the effective temperature and the surface gravity of the M stars
in our identified WDMS binary sample have been obtained. We can
derive the spectral type of the M dwarf in each WDMS binary
according to an empirical ${\rm Sp}-T_{\rm eff}$ relation proposed
by \citet{reb07}. The distribution of the M dwarf spectral types
in our WDMS binary sample (including a DA/DB WD component) is
displayed in Fig. 4, in which the solid line represents the all
DA/DB WDs in WDMS binaries identified by us from SDSS DR9 and DR10
and the shaded zone represents the new DA/DB WDs in WDMS binaries
found from DR10, together with the DB WDs in WDMS binaries found
by us from SDSS DR9. It is found in Fig. 4 that the spectral types
of the majority of M dwarf components in WDMS binaries are located
in a region from M3 to M6 in both cases. This is similar to the
results found by \citet{Heller09,Liu12} and
\citet{reb07,reb12,Reb13}.

\begin{table*}
\tiny
 Table~1.\hspace{4pt} Object name, coordinates (in degrees),
and SDSS $ugriz$, 2MASS $JHK$, UKIDSS $yjhk$ and WISE $w_{1,2}$
magnitudes of the 468 WDMS binaries identified from SDSS DR9 and
DR10. The
complete table is available in the electronic edition of the paper.\\
\begin{minipage}{17cm}
\begin{tabular}{l|cccccccccccccccc}
\hline\hline\\
SDSS J             &  RA &DEC&  u    &   g   &    r  & i    &   z  & J    &  H   &K   & y     & j    & h      &   k &   w1 &  w2 \\
\hline
000756.04+055723.5 & 1.98349 & 5.95653 &21.05& 20.52& 20.50& 20.04& 19.49  &-- &  -- &  -- & 18.65 &18.21 & -- & -- & 16.80& 16.62\\
001707.14+040145.3 &  4.27974 &  4.02926& 21.65& 21.02 &21.15& 20.71& 20.31& -- &  -- &-- & 19.34& 18.77& 18.25 &17.81& 17.65& 16.46\\
001744.95+001203.6&  4.43731 & 0.20100& 20.40& 19.92& 20.08& 20.63& 19.35& 13.97& 13.38& 13.14& 14.42& 13.98&-- & 13.15& 13.03& 12.99\\
002438.47$-$024024.7&  6.16029 & -2.67352& 21.47& 20.89& 20.79& 20.20& 19.54&  -- &-- &--&-- &-- &--&-- & 17.07& 16.57\\
003509.76+023153.5 & 8.79068 &2.53153& 21.43& 20.93& 21.10& 20.94& 20.43&  -- &-- &-- &19.54& 19.27&18.60& 18.28 &-- &-- \\
003830.93+081457.2& 9.62887& 8.24921& 21.00& 20.60& 20.53& 20.49& 20.20&  -- &-- &-- &20.01 &--&-- &-- & 16.51& 16.23\\
003955.54+104702.9& 9.98143&10.78413&20.72 &20.26& 20.36& 20.15& 19.77&  --   &-- &-- &-- &-- &-- &-- &-- &-- \\
004212.74+004220.6&10.55310 & 0.70573& 21.45& 20.94& 20.72& 19.78 &19.03&  -- &-- &-- &-- &-- & 17.31& 17.08& 16.87& 16.72\\
004435.57$-$000036.7& 11.14822 & -0.01020& 19.13& 17.68& 16.69& 16.22& 15.94& 14.78& 14.13 &14.01 &-- &-- &-- &-- & 13.99& 13.96\\
004643.54+084227.5& 11.68143& 8.70763& 19.48& 19.37& 19.48& 19.40& 19.02&  -- &-- &-- & 18.14& 17.59& 16.95 &16.68& 16.49& 15.95\\
004827.34+075939.8& 12.11392& 7.99438& 19.47& 19.31& 19.39& 18.73& 18.10& 16.88& 15.81&15.67& 17.25& 16.72& 16.25& 16.01& 15.85& 15.71\\
005308.36+085631.6& 13.28483& 8.94212& 21.58& 21.00& 20.80& 20.25& 19.52&  --&--&--& 18.75& 18.07& 17.78& 17.46& 17.45& 16.38\\
005443.84+070921.2& 13.68265& 7.15590&18.20&18.41& 18.72& 18.83& 19.59& 16.78 &15.79& 17.11& 17.34& 16.77& 16.20& 15.95 &15.70& 15.50\\
005752.57+035851.6& 14.46905& 3.98101& 20.85& 20.42& 20.58& 20.36& 19.99&  --&--&--& 19.15& 18.73 &18.19 &17.88&-- &-- \\
005827.26+005642.6& 14.61359& 0.94516& 23.04& 21.21& 19.96& 18.50& 17.56& 16.04& 15.75& 14.97&--& 16.08& 15.51& 15.23& 14.77& 14.56\\
005906.65$-$003802.2& 14.77772 & -0.63394& 21.98& 20.72& 19.61& 18.25& 17.38& 16.00& 15.64& 15.00&--&--&--&--& 14.97& 14.73\\
010055.32+074758.9& 15.23052& 7.79970& 19.23& 18.98& 19.21& 19.07& 18.65&  -- &--&-- & 17.76& 17.22& 16.73& 16.42&-- &-- \\
\hline
\end{tabular}
\end{minipage}
\end{table*}

\subsection{WDMS binaries with several SDSS spectra}

In this work, we have identified 121 WDMS binaries with several
SDSS spectra observed in different times, They are listed in Table
3. Two (SDSS J220451.62+113230.8 and SDSS J111544.56+425822.4) of
these WDMS binaries are newly identified by us from SDSS DR10,
their spectra are displayed in Fig. 5 and Fig. 6, respectively. As
seen from Fig. 5, SDSS J220451.62+113230.8 have been observed
three times by SDSS spectral survey, however its spectra observed
in different times do not show any obvious variation in the radial
velocities of its two components. This suggests that it might be a
PCEB with a low orbital inclination or it is a WDMS binary with a
wide separation and a long orbital period so that the variation in
its radial velocity can not be found by us in a short period.
Meanwhile, one (SDSS J095043.96+391541.7) of these WDMS binaries
was found by us from SDSS DR9 \citep{Li14}, its spectra are shown
in the top panel of Fig. 6. Another two WDMS binaries [SDSS
J131751.72+673158.4 and SDSS J142947.62-010606.9 listed in
\citet{Reb13} and \citet{Sil06}, respectively] are also plotted in
Fig. 6 since their spectra shows the same properties as those of
SDSS J111544.56+425822.4 and SDSS J095043.96+391541.7 recently
identified from SDSS by us.

It is found in Fig. 6 that the SDSS observed spectra of four WDMS
binaries (SDSS J095043.96+391541.7, SDSS J111544.56+425822.4, SDSS
J143947.62-010606.9, and SDSS J131751.72+673158.4) share a similar
characteristics, i.e. their SDSS spectra exhibit Hydrogen emission
in all observable Balmer series lines in addition to He I
emission. This phenomenon was primarily found in SDSS
J143947.62-010606.9 by \citet{Sil06}, who suggest this is caused
by the photoionization and recombination induced by irradiation of
the M dwarfs in these WDMS binaries rather than the mass accretion
as that in CVs, since the emission line width ($\la 20$ angstroms)
in these WDMS binaries is much narrower than that ($\sim 30-40$
angstroms) in CVs. As seen from Fig. 6, the WDs in such WDMS
binaries usually have a very high effective temperature based on
their spectral profile, and therefore they are very young WD
stars. In addition, it is found from the sub-plot in the upper
right corner of each panel of Fig. 6 that the components in three
binaries exhibit spectral line shift. In particular for SDSS
J131751.72+673158.4, its two SDSS spectra span an interval of only
one day (i.e. MJD55622 and MJD55623). This suggests that such WDMS
binaries should be the young PCEBs with a short period. Although
SDSS J143947.62-010606.0 does not show any variation in the radial
velocity of its WD component, it might be the very young
shoet-period PCEBs with a very low inclination. We would obtain
the orbital periods on the basis of their radial velocity
observations in the future.

\begin{table*}
\begin{footnotesize}
Table~2.\hspace{4pt} Stellar parameters for 468 WDMS binaries identified from SDSS DR9 and DR10. The complete table is available in the electronic edition of the paper.\\
\begin{minipage}{17cm}
\tiny
\begin{tabular}{l|cccccccccccccccccc}
\hline\hline\\
 SDSS J& Type &PLT&MJD&FIB&$T_{\rm eff}^{\rm WD}$&${\rm log}g_{_{\rm WD}}$&$T_{\rm eff}^{\rm sec}$&${\rm log}g_{\rm sec}$&$d_{_{\rm WD}}$&$M_{\rm WD}$      &
$r_{_{\rm WD}}$   &$d_{\rm M}$&$M_{\rm sec}$    &$r_{\rm sec}$     &$C$& ${\rm log}Age $&$\chi_{\rm red,min}^{2}$& note\\
       &      &   &   &   &(K)                   &    (${\rm cm/s^{2}}$)         &   (K)    &    (${\rm cm/s^{2}}$)   & (pc)          &($M_{\rm \odot}$) &
($R_{\rm \odot}$) &(pc)         &($M_{\rm \odot}$)&($R_{\rm \odot}$) &   &(yrs) & &\\
 \hline
000756.04+055723.5& DA+dM& 4416&55828& 502& 18000& 8.250&3020& 4.0&  771.11  &0.778&0.011&577.95& 0.26&0.26&0.20&8.25&2.13& N\\
001707.14+040145.3& DA+dM& 4299& 55827& 686& --&  --&  -- & -- & --&     --     &--& -- & -- & -- & -- & -- & -- & N\\
001744.95+001203.6& DA/dM &4219& 55480& 618& 17000& 8.000&3843& 5.0&  653.36 & 0.620 & 0.013 & 732.35 & 0.47 & 0.49& -0.08& 8.15 & 2.92& N\\
002438.48-024024.7& DA/dM& 4367& 55566& 278& 19000& 8.250&3020& 4.0& 1027.19&0.780& 0.011& 620.68& 0.26 &0.26 & 0.35& 8.18 & 1.31& L\\
003509.76+023153.5& DA/dM& 4304& 55506& 520& 25000& 8.750&2281.0& 3.5&  899.07&1.073& 0.007 &147.75&  0.13& 0.12 & 1.01 & 8.20 & 6.00& L\\
003830.93+081457.2& DA/dM& 4542& 55833 &444& 25000& 7.750&3843& 5.5& 1643.46& 0.529& 0.016& 2554.01 & 0.47&0.49 & -0.31& 7.27& 1.72& N\\
003955.54+104702.9& DA/dM& 5656& 55940& 110& 18000& 8.250& 2856& 3.5&  688.67& 0.778& 0.011& 522.33& 0.20& 0.20 & 0.19&  8.25 &0.65& N\\
004212.74+004220.6& DA/dM& 3589 &55186& 680& 25000& 8.750& 3185& 4.5&  904.01 & 1.073 & 0.007 & 504.30 & 0.32 & 0.33 &0.40 & 8.20 & 2.23& L\\
004435.57-000036.7& WD/dM& 3588& 55184 &  34 &  -- &  -- &   -- &  -- &   --  & --  & --  & --  & -- & -- &  -- & -- &  -- & N\\
004643.54+084227.5& DB/dM& 4544& 55855 & 458 & 8500& 7.000& 3020& 3.5&  464.75 & 0.189& 0.023 & 370.18&0.26 & 0.26 & 0.16 & 8.71 & 1.75& N\\
004827.34+075939.8& DA/dM &4543& 55888 & 890& 25000& 8.250& 3185& 5.0&  614.92 & 0.792 & 0.011& 335.26 & 0.32 & 0.33 & 0.42 & 7.76& 1.55& N\\
005308.36+085631.6& DA+dM& 4546& 55835 & 590 &  -- &  -- &  -- &  -- &  -- & --  & --  & --  & -- & -- &  -- & -- &  -- & N\\
005443.84+070921.2& DA/dM& 4546& 55835 & 304& 25000& 8.250& 3843& 5.5&  434.82 & 0.792& 0.011& 1201.24 & 0.47 & 0.49 & -0.66& 7.76 & 4.55& N\\
005752.57+035851.6& DA/dM& 4307& 55531& 858& 25000& 9.000& 3185& 4.0 & 499.27 & 1.203 & 0.006& 749.06& 0.32& 0.33& -0.28& 8.40& 2.29& L\\
010055.32+074758.9& DA/dM& 4546& 55835& 996& 19000& 8.250& 3185& 4.5&  383.30&0.780 & 0.011 & 445.49 & 0.32 & 0.33 & -0.11 & 8.18 &7.47& N\\
 \hline
\end{tabular}
\end{minipage}
\end{footnotesize}\\
{Notes in Table 2: (1) L: WDMS binaries identified by Li (2014);
(2) N: New WDMS binaries identified from DR10}
\end{table*}

\begin{table*}
Table~3.\hspace{4pt} Object name, coordinates (in degrees), plates, MJD, and Fibres of
121 WDMS binaries observed by SDSS several times. The complete table is available in the electronic edition of the paper.\\
\begin{minipage}{16cm}
\begin{tabular}{l|cccccc}
\hline\hline\\
SDSS J  & ${\rm RA}$ & ${\rm DEC}$ & PL-MJD-Fib& PL-MJD-Fib&
PL-MJD-Fib&PL-MJD-Fib\\ \hline
$001749.25-000955.4$ &   4.45519 & -0.16539 &0389-51795-0112 & 0687-52518-0109&4218-55479-0064&               \\
002801.68+002137.7 &   7.00701 &  0.36047 &0689-52262-0424 & 3586-55181-0570&               &               \\
$003628.07-003125.0$ &   9.11695 & -0.52360 &0690-52261-0222 & 3586-55181-0072&0689-52262-0022&               \\
$005457.61-002517.2$ &  13.74004 & -0.42145 &0394-51913-0110 & 4224-55481-0228&0394-51812-0118&0394-51876-0109\\
010345.56+003746.7 &  15.93985 &  0.62965 &1498-52914-0466 & 3735-55209-0723&1083-52520-0638&1497-52886-0615\\
012259.56+154253.9 &  20.74817 & 15.71496 &0424-51893-0404 & 5140-55836-0604&               &               \\
$014745.03-004911.1$ &  26.93761 & -0.81976 &0699-52202-0094 & 3606-55182-0484&               &               \\
$015225.39-005808.6$ &  28.10579 & -0.96906 &1504-52940-0083 & 3606-55182-0294&               &               \\
$023435.58-004818.4$ &  38.64824 & -0.80512 &0706-52199-0296 & 3745-55234-0340&0705-52200-0017&               \\
$023804.40-000545.7$ &  39.51831 & -0.09602 &0706-52199-0190 & 3745-55234-0200&               &               \\
024505.48+002804.6 &  41.27282 &  0.46794 &0707-52177-0471 & 3651-55247-0805&               &               \\
$025123.34-011314.4$ &  42.84724 & -1.22068 &0708-52175-0243 & 4241-55450-0124&               &               \\
\hline
\end{tabular}
\end{minipage}
\end{table*}

\begin{figure}
\centerline{\psfig{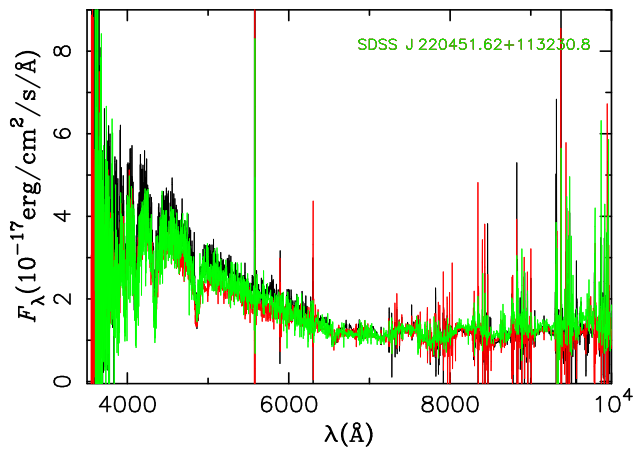}}
\caption{Shown is the spectra of a new WMDMS binary with three
different SDSS spectra.}
 \label{fig5}
\end{figure}

\subsection{Distance distribution for WDMS binaries}

As mentioned sect. 3.1, the two independent distance estimates for
each DA/DB WD and its M dwarf companion from the earth have
determined according to the flux scale factors obtained from the
best fitting the SDSS observed spectra of WDMS binaries. We
compare the distances of the white dwarfs with those of their
M-dwarf companions in Fig. 7. In Fig. 7, the symbol pluses
represent 396 analyzed WDMS binaries identified from SDSS DR9 and
DR10, the solid dots represent the new WDMS binaries identified
from DR10, together with those containing a DB WD component which
the physical parameters are newly derived in this work, the solid
line denotes the case with $d_{\rm WD}=d_{\rm M}$, the two dashed
lines span a tolerance fan for $C=0.5$ as \citet{Li14}. As seen
from Fig. 7 and Table 2, about 80 percent of the WDMS binaries
with a distance discrepancy coefficient $C$ \citep{Heller09}
determined base on Eq. (3) are in a range between $-0.5$ and $0.5$
in both cases. This suggests that about 80 percent of the WDMS
binaries with a DA/DB white dwarf component have $d_{\rm WD}\simeq
d_{\rm sec}$, implying that about 20 per cent of these binary
systems show a significant difference between the two distance
estimates for the WDs and M-dwarfs. This is similar to the results
derived by \citet{reb07}, \citet{Heller09}, \citet{Liu12} and
\citet{Li14}.

\begin{figure}
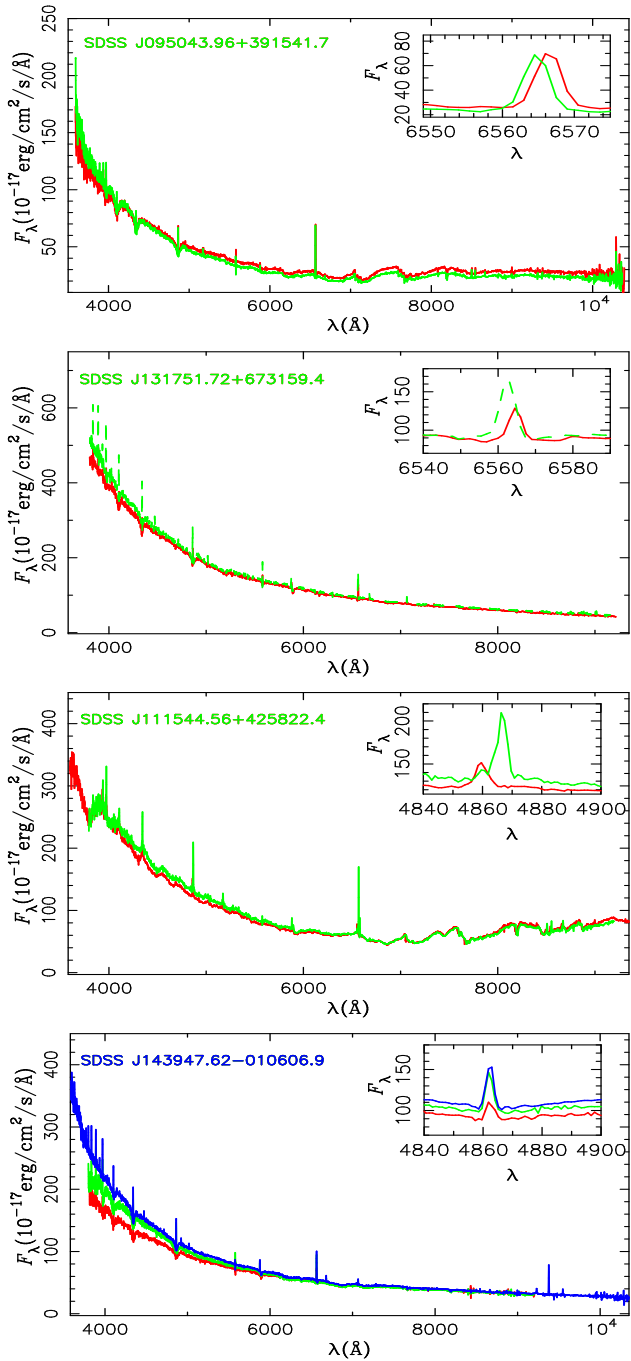

\centerline{\psfig{figure=Fig11.ps,width=8.5cm,height=4.5cm,bbllx=527pt,bblly=109pt,bburx=179pt,bbury=628pt,angle=270}}
\centerline{\psfig{figure=Fig12.ps,width=8.5cm,height=4.5cm,bbllx=525pt,bblly=109pt,bburx=187pt,bbury=628pt,angle=270}}
\centerline{\psfig{figure=Fig13.ps,width=8.5cm,height=4.5cm,bbllx=525pt,bblly=109pt,bburx=187pt,bbury=628pt,angle=270}}
\centerline{\psfig{figure=Fig14.ps,width=8.5cm,height=4.5cm,bbllx=525pt,bblly=107pt,bburx=187pt,bbury=628pt,angle=270}}
\caption{Shown is the spectra of four specific WDMS binaries with
hydrogen line emission and Helium line emission. Top panel: the
SDSS spectra of SDSS J095043.96+391541.7. The second panel: the
spectra of SDSS J131751.72+673159.4 \citep[listed in][]{Reb13}.
The third panel: the spectra of SDSS J111544.56+425822.4. Bottom
panel: The spectra of SDSS J143947.62-010606.9 \citep{Sil06}.}
\label{fig6}
\end{figure}

\section{Discussion and conclusion}

In this work, we have identified 2198 new WDMS binaries from SDSS
DR10 based on a color-selection criteria \citep{Li14} and a
color-selection criteria for searching WDs from SDSS or LAMOST
(Gong et al. 2014, in preparation). 1889 of which have been
discovered by previous investigators
\citep[e.g.][]{Ray03,Sil07,Heller09,Reb13,Liu12,Li14}. Therefore,
309 WDMS binaries are newly identified by us from DR10, these
binaries include 220 DA+MS binaries, 32 DB+MS binaries, and 57
WDMS binaries containing a WD component with the other spectral
types of WDs. \citet{Kepler14} reports 180 WDMS binaries
identified from SDSS DR10, 120 of which have been discovered by
\citet{Li14} and \citet{Reb13}. Therefore, about 60 WDMS binaries
recently found by \citet{Kepler14} are probably in 309 WDMS
binaries discoverd by us independently.

\begin{figure}
\centerline{\psfig{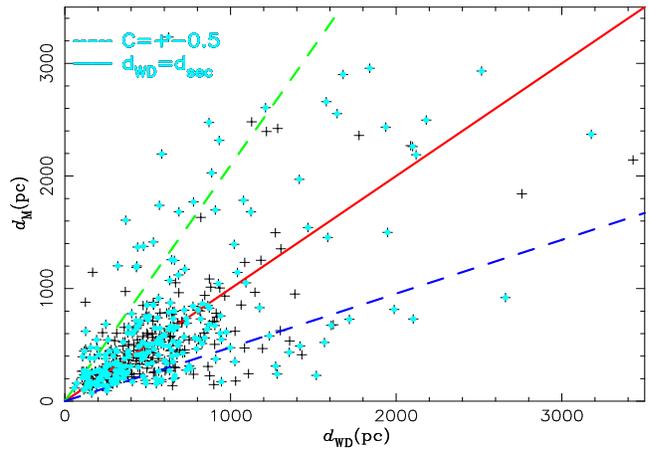}}
\caption{The distribution of the distances of the two components
in WDMS binaries identified by us from SDSS. The solid line
denotes the case with $d_{\rm WD}=d_{\rm sec}$, the blue dashed
line denotes the case with $C=+0.5$ and the green dashed line
represents the case with $C=-0.5$. The pluses represent all WDMS
binaries \citep[including new WDMS binaries and those found
by][]{Li14}, the solid dots represent those newly found from DR10,
together with those containing a DB WD component identified by Li
et al. (2014).}
\label{fig7}
\end{figure}

Based on a $\chi^2$ minimization technique originally developed by
\citet{Heller09} and \citet{reb07}, we have determined the
independent stellar parameters (effective temperature, surface
gravitation, mass, radius and distance) of the two components in
237 new WDMS binaries with a DA/DB WD component. The other 15 new
WDMS binaries with a DA WD component have a very low SN, the
stellar parameters for their components cannot be precisely
determined in this work. The fraction of the WDs with a mass
larger than $\sim 0.9M_{\rm\odot}$ in \citet{Li14} is slightly
higher than those in the previous investigators' works
\citep[e.g.][]{Sil06,reb07,Reb13}. One possible explanation for
this behaviour is that the SN of the observed spectra has a
significant influence on the determination of the metallicity of
the M dwarfs and the surface gravity of two components, and maybe
on the determination of the masses of the WD stars in WDMS
binaries. We analyze the spectra of these WDMS binaries again
based on an assumption that the metallicity of the M dwarfs in
these binaries is the solar one. In fact, all of these binaries
are in the neighborhood of the sun, and the atmosphere of the M
dwarfs in WDMS binaries should not be seriously polluted by the
progenitors of the WDs in these systems because the timescale of
the CE phase is very short and the M dwarfs usually have a deep
surface convective zone. Another possible explanation for this
behaviour is that the so-called high-${\rm log}g$ problem is
presented in 1D model atmosphere of DA WDs \citep{Tremblay13}. In
this work, the values of the effective temperature and surface
gravity of the DA WDs (with $T_{\rm eff}\la 14,000$ K) determined
in 1D models are revised to those in 3D model atmosphere for the
DA WDs \citep{Tremblay13}. Then, the average mass of the WDs in
new WDMS binaries and in all WDMS ones identified by us is derived
to be of $0.632 M_{\rm \odot}$ and $0.692M_{\rm \odot}$,
respectively. This is similar to the results recently derived by
\citet{Kepler14}.

The stellar parameter distributions have used to probe the general
properties of the WDMS binaries, it is found in Fig. 2 that the
effective temperature of the WDs in the majority of our WDMS
binaries is lower than 25,000K, the surface gravity of the WDs in
WDMS binaries peaks at ${\rm log}g_{\rm WD}\sim 8.0$, and the mass
of them peaks at $\sim 0.5 M_{\rm \odot}$. As shown in Fig. 3, the
cooling age of the WDs peaks at ${\rm log}{\rm Age_{cool}}\sim
8.6$, the mean cooling age of the WDs is of $\sim 2.0\times10^{8}$
yrs, and the young WDs are very rare. Meanwhile, the spectral
types of M-dwarfs are derived from their effective tempwerature
through an empirical ${\rm Sp}-T_{\rm eff}$ relation
\citep{reb07}. As seen from Fig. 4, the spectral types of most
WDMS binaries are located in a region between M3 and M6. These
results are consistent with the results derived by the previous
investigators
\citep[e.g.][]{Sil06,Heller09,Liu12,reb07,reb12,Reb13}.

121 WDMS binaries are found to be observed by SDSS spectral survey
several times. In these binaries, at leat three such binaries
(SDSS J220451.62+113230.8, SDSS J095043.96+391541.7 and SDSS
J111544.56+425822.4) are recently discovered by us from SDSS DR9
and DR10, respectively. Four WDMS binaries, J095043.96+391541.7,
SDSS J111544.56+425822.4, SDSS J143947.62-010606.9 \citep{Sil06},
and SDSS J131751.72+673159.4 \citep[listed in][]{Reb13} show
hydrogen emission in all observable Balmer series lines in
addition He I emission (see Fig. 6). A possible explanation for
this behaviour is photoionization and recombination of M dwarfs
caused by the irradiation of M dwarfs due to their very hot WD
companions. It is found in Fig. 6 that the WDs in these WDMS
binaries usually have a very high effective temperature, and
therefore they should be very young WDs according to their
spectral profile (such as SDSS J095043.96+391541.7 with $T_{\rm
eff}=60,000$ K and ${\rm log}Age=5.28$ and SDSS
J111544.56+425822.4 with $T_{\rm eff}=40,000$ K and ${\rm
log}Age=7.71$). Also as seen from Fig. 6, three of these binaries
show the shift of ${\rm H_{\alpha}}$ or ${\rm H_{\beta}}$ emission
line, implying that the components of these binary systems exhibit
the variation in their radial velocities, especially for the
system SDSS J095043.96+391541.7, its two SDSS spectra only cover a
time interval of one day (MJD55622 and MJD55623). This suggests
that four such WDMS binaries should be PCEBs with a short orbital
period, and the system SDSS J143947.62-010606.9 should be a PCEBs
with very low orbital inclination. Since these young PCEBs are not
polluted not only by mass transfer but also by angular momentum
loss, they have a special significance for limiting the results of
CE phase of the binary evolution. We will monitor the variation in
the radial velocities of these binaries to obtain their orbital
periods.

Two independent distances for two components in WDMS binaries are
derived on the basis of two flux-scaling factors described in
\citet{Heller09}. It is found in Fig. 7 that 80 per cent of the
WDMS binaries with a DA/DB white dwarf and an M dwarf satisfy
$d_{\rm WD}\simeq d_{\rm sec}$ (i.e. they locate in a range with
$\mid C\mid <0.5$). Approximately one-fifth of WDMS binary systems
suffer a significant difference between $d_{\rm WD}$ and $d_{\rm
sec}$. This is similar to the results derived by \citet{Liu12} and
\citet{reb07}. The significant difference between $d_{\rm WD}$ and
$d_{\rm sec}$ might be a result of the ${\rm H}_{\rm \alpha}$
emission and/or other types of Balmer emission (such as SDSS
J143012.51+250437.6, SDSS J213019.79+061204.6, and SDSS
J215713.29-001958.8) owing to the stellar activity of the M dwarfs
in these WDMS binaries, which might cause a too early spectral
type to be determined for the M dwarfs in WDMS binaries
\citep{Heller09,reb07,Sil06}. Meanwhile, another explanation for
this behaviour is the expansion in the radii of M dwarfs
irradiated by their close hotter WD primary
\citep{Sil06,Heller09}, such as in SDSS J095043.96+391541.7 and
SDSS J 111544.56+425822.4 mentioned above.

\section*{ACKNOWLEDGEMENTS}

This project was partly supported by the Chinese Natural Science
Foundations (Grant Nos. 11373063, 11073049, 11273053, 11033008,
10821026, 11390374 and 2007CB15406) and Yunnan Foundation (Grant
No. 2011CI053), and by the Chinese Academy of Sciences
(KJCX2-YW-T24).

\bsp

\label{lastpage}

\end{document}